
\input phyzzx
\REF\original{A. A. Belavin, in  $\ Quantum \ String \ Theory$,
Volume 31, Springer-Verlag, Berlin, Heidelberg, 1988 page 132.}
\REF\Alek{A. Alekseev and Shatashvilli, Nucl. Phys B323 (1989)
719.}
\REF\Sok{V. Drinfeld and V. Sokolov, Journal of Soviet Mathematics
30, (1984) 1975.}
\REF\oog{M. Bershadsky and H. Ooguri, Comm. Math.
Phys. 2 (1989) 49.}
\REF\superoog{ M. Bershadsky and H. Ooguri, Phys. Lett, 229B
(1989) 374.}
\REF\diff{ A. M. Polyakov, Int. J. Mod. Phys. A5(1990)
833.}\REF\poly{A.M. Polyakov, Mod. Phys. Lett. A2 (1987) 893;
\hfill\break V.G. Knizhnik, A.M. Polyakov and A.B. Zamolodchikov,
Mod. Phys.  Lett.  A3 (1988) 819; \hfill\break
A.M. Polyakov and A.B. Zamolodchikov, Mod.  Phys. Lett. A3
(1988)1213;\hfill\break
Lectures given by Polyakov at Les Houches summer school on Fields,
Strings and Critical Phenomena (1988).}\hfill\break
\REF\birk{ W. A. Sabra, Nucl. Phys B375 (1992) 82.}
\REF\bbirk{ W. A. Sabra,  Restricted gauge invariance, N=2
Coadjoint orbits \hfill\break
 and  N=2 quantum
supergravity, preprint, BIR/PH/92-1.}
\REF\waki{M. Wakimoto, Comm. Math. Phys. 104 (1986) 605.}
\REF\unpublished{ A.B. Zamolodchikov, unpublished}
\REF\Novi{S. P. Novikov, Sov. Math. Dokl. 24 (1981) 222.}
\REF\WIT{E. Witten, Comm. Math. Phys. 92 (1984) 455.}
\REF\composition{A.M. Polyakov and P.B. Wiegmann, Phys. Lett. 141B
(1984) 233.}
\REF\superkuramoto{T. Kuramoto,  Quantum Hamiltonian Reduction Of
Super Kac-Moody Algebra II, preprint, YNUE-PH-92-01.}
\REF\fuchs{B. L. Feigin
and D. B. Fuchs,  Funct. Anal. and Appl.  16 (1982) 114;  17 (1983)
241.} \REF\Peterson{P. di Vecchia, V. G. Knizhnik, J. L. Peterson
and P. Rossi,  Nucl. Phys B253 (1985) 701.}
\REF\FUCHS{J. Fuchs,  Nucl. Phys B286 (1986)
455, B318 (1989) 631.}
\REF\B{V. G. Knizhnik, Theor. Math. Phys. 66 (1986) 68; M.
Bershadsky, Phys. Lett. 174B (1986) 285.} \date={April 1992}
\rightline {April, 1992}
\rightline { BIR/PH/92-2.}
\title {CLASSICAL HAMILTONIAN REDUCTION AND SUPERCONFORMAL ALGEBRAS}
\author{W. A. Sabra\foot{email ubap734@uk.ac.bbk.cu}}
\address {Physics Department \break Birkbeck College,\break Malet
Street\break London WC1E 7HX}
\abstract{
The Polyakov's \lq\lq soldering procedure" which shows how
two-dimensional diffeomorphisms can be obtained from SL(2,R) gauge
transformations is
discussed using the free-field representation of SL(2,R) current
algebra. Using this formalism, the relation of Polyakov's method to
that of the Hamiltonian reduction becomes transparent. This
discussion is then generalised to N=1 superdiffeomorphisms which can
be obtained from N=1 super Osp(1,2) gauge transformations.
It is also demonstrated that the phase space of the
Osp(2,2) supercurrent algebra represented by free superfields
is connected to the classical phase space of
N=2 superconformal algebra via
Hamiltonian reduction.} \endpage
\chapter{Introduction} In recent
years, the Hamiltonian reduction of the phase space of the
Wess-Zumino-Novikov-Witten (WZNW) models with the gauge group
SL(2,R) and its graded extensions [\original-\diff] provided more
understanding not only of the structure of conformal and
superconformal algebras but also of the origin of non-compact
current algebra symmetries in the chiral gauge formulation of
induced two dimensional gravity and supergravity theories [\poly].
Classically, the Hamiltonian reduction [\Sok, \oog] can be described
as follows. A constraint is imposed on the affine currents
describing the phase space of the WZNW model. The constrained phase
space modulo gauge transformations, generated by the residual
symmetry preserving the constraints, is the phase space of the
reduced theory.

In [\oog] by imposing a certain constraint on the
SL(2,R) Kac-Moody current algebra,
an irreducible representation of the Virasoro algebra was obtained
from an irreducible representation of the SL(2,R) Kac-Moody current
algebra using the BRST formalism. This method is the quantum
analogue of the Hamiltonian reduction.
The hidden SL(2,R) current algebra symmetry in the
Virasoro algebra manifests itself in the chiral gauge
formulation of induced two dimensional quantum gravity [\poly]. The
results of [\oog] were extended to the supersymmetric cases
in [\superoog], in which the N=1 and N=2 superconformal algebras
were obtained as the Hamiltonian reduction of the bosonic Kac-Moody
current algebras of Osp(1,2) and Osp(2,2) respectively. In this
construction,  extra fermionic degrees of freedom have to be
introduced in order to get a consistent algebra of constraints.

In refs. [\birk,\bbirk], following the
method of Polyakov in which the conformal transformation of
the stress energy tensor defining the phase space of Virasoro
algebra was derived from the restricted SL(2,R) gauge
transformations, the N=1 and N=2 superconformal transformations were
derived as restricted  N=1 Osp(1,2) and Osp(2,2) supergauge
transformations respectively.  The method of Polyakov is simply a
Hamiltonian reduction in disguise. This is best explained by using
free fields (Wakimoto fields) [\waki,\unpublished] to describe the
phase space of SL(2,R) WZNW model. Introduce a scalar field $\phi$
and a pair of bosonic ghosts $(\beta,\gamma)$ of
dimensions $(1,0).$
The Poisson brackets of these fields are given by
\footnote *{we suppress the space-time
indices of the fields and parametrize the two dimensional space
time with coordinates $(x, t)$ by $z=t+x$ and $\bar z=t-x$.}
$$\eqalign{\{\beta(z_1),\gamma(z_2)\}_{P.B}=&-\delta(z_1-z_2),
\cr\{\partial_{z_1}\phi(z_1),\phi(z_2)\}_{P.B}=&-{1\over2k}\delta(z_1-z_2).\cr}
\eqn\poi$$The SL(2,R) conserved classical currents can be
written as
$$\eqalign
{J^+=&k\partial_z\gamma+2k\gamma\partial_z\phi-\gamma^2\beta,\cr
J^-=&\beta,\cr
J^0=&-\beta\gamma+k\partial_z\phi,\cr}\eqn\ali$$ where $k$ is the
level of the SL(2,R) current algebra and $\{0,+,-\}$ are the
indices of the SL(2,R) group. One way to construct the
currents in terms of the free fields is to set $J^{-}=\beta$
and write the most general form of $J^0$ and $J^+$ in
terms of the free fields $\beta, \gamma$ and $\partial_z\phi$, which
is then fixed by demanding that the Poisson brackets of the currents
satisfy the classical SL(2,R) current algebra [\unpublished]. The
currents can also be derived from the SL(2,R) WZNW model as follows.
One can parametrize the group element $g$ of SL(2,R) by the Gauss
product [\Alek]
$$g=\pmatrix{1&0\cr \gamma&1\cr}\pmatrix{\lambda&0\cr 0&\lambda^{-
1}\cr}\pmatrix{1&F\cr 0&1\cr}.\eqn\tr$$
Then the SL(2,R) WZNW action [\Novi,\WIT,\composition] can be
written as $$S\sim k\int d^2z
\Big({1\over\lambda^2}\partial_z\lambda\partial_{\bar
z}\lambda+\lambda^2\partial_zF\partial_{\bar
z}\gamma\Big).\eqn\sun$$
The right-moving conserved currents derived
from the action \sun\  are given by
$$\eqalign{{J}=\pmatrix{J^0&J^-\cr
J^+&-J^0\cr}&=k\partial_zgg^{-1}\cr &=
k\pmatrix{{\partial_z\lambda/\lambda}-\gamma\lambda^2\partial_z
F&\lambda^2\partial_z F\cr
\partial_z\gamma+2\gamma{\partial_z\lambda/\lambda}
-\gamma^2\lambda^2\partial_z
F&-{\partial_z\lambda/\lambda}+\gamma\lambda^2\partial_z
F\cr}.\cr}\eqn\yan$$ By setting $$k\lambda^2\partial_z F=\beta,
\qquad \lambda=e^\phi, \eqn\yann$$  we get the expressions of the
currents in \ali\ and the action \sun\ becomes an action of a free
scalar field and a bosonic ghost system,
$$S\sim \int d^2z\Big(k\partial_z\phi\partial_{\bar
z}\phi+\beta\partial_{\bar z}\gamma\Big).\eqn\frust$$ Polyakov's
partial gauge fixing  [\diff] is given as $$J^-=k,\qquad J^0=0,
\qquad J^+=-T.\eqn\desert$$  Solving \desert\ in
terms of the free fields gives
$$T=-k(\partial_z^2\phi+(\partial_z\phi)^2),\eqn\pinhead$$ which is
recognized as the classical Feign-Fuchs equation [\fuchs] for the
spin-2 generator describing the phase space of Virasoro algebra.
The Poisson bracket of $T$ is calculated using \poi\ and is given by
$$\{T(z_1),T(z_2)\}_{\rm
P.B}={k\over2}\partial_{z_2}^3\delta(z_1-z_2)+
2T(z_2)\partial_{z_2}\delta(z_1-z_2)+\partial_{z_2}
T(z_2)\delta(z_1-z_2).\eqn\nest$$ The reason why $J^+$ becomes a
spin-2 field was explained in [\diff] as a result of the \lq\lq
soldering" of the isospin space and the two dimensional space time
in the background $J^-=k$.  In terms of the free fields this can be
explained as a result of the fact that after the partial gauge
fixing, the field $\gamma$ originally of spin-0 becomes a spin-1
field ($\gamma=\partial_z\phi$). In the classical Hamiltonian
reduction method [\Alek,\oog], the constraint $J^-=k$ is first
imposed. Then the constrained phase space has a one-parameter
residual symmetry generated by the Borel subgroup of SL(2,R), this
residual symmetry can be used to gauge away a further degree of
freedom giving a reduced  theory with one degree of freedom.  An
element of the space of the residual symmetry that preserves the
constraint $J^-=k$ is given by  $$u=\pmatrix{1&0\cr
a&1\cr},\eqn\bo$$ where $a$ is a gauge parameter. Under the finite
residual gauge transformation $$g\rightarrow ug,\eqn\tr$$ The
currents transform as $${J}\rightarrow
uJu^{-1}+k\partial_zuu^{-1}.\eqn\lll$$ Exploiting the gauge symmetry
\lll\ and  selecting the gauge parameter as (Drinfeld-Sokolov
gauge) $$a={1\over k}{J}^0,\eqn\dual$$ one can put ${J}$ into the
form $${ J}_{DS}=\pmatrix{0&k\cr -T&0\cr},\eqn\safia$$ where $T$, the
coordinate of the reduced phase space is given by Eq.\pinhead. This
demonstrates that Polyakov's partial gauge fixing
is equivalent to the Drinfeld-Sokolov gauge
in the Hamiltonian reduction method.

It has to be mentioned that there also exists another
gauge choice, the so called diagonal
gauge in which the reduced space takes the form
$${\bf J}_{diagonal}=k\pmatrix{\partial_z\phi&1\cr
0&-\partial_z\phi\cr}.\eqn\dia$$ The connection between
the coordinates of the reduced space in the two different
gauges can be explained as follows. Consider the
system of linear differential equations
$$(k\partial_z-J)\pmatrix{v_1\cr v_2}=0,\eqn\coffee$$ where $v_1$
and $v_2$ are the components of a two-dimensional vector. In the
above system of differential equations one can eliminate the
component $v_2$ and obtain a gauge invariant second order
differential equation for $v_1$.  Computing \coffee\ in the diagonal
gauge gives  $$\eqalign{k\partial_zv_1-k\partial_z\phi
v_1-kv_2=&0,\cr k\partial_zv_2+k\partial_z\phi
v_2=0,\cr}\eqn\champs$$ after eliminating $v_2$ in \champs\  we get
$$k\partial_z^2v_1-k\partial_z\phi\partial_z\phi
v_1-k\partial^2_z\phi v_1=0\eqn\vogue$$ The reduced space of
Drinfeld-Sokolov gauge can be deduced from that of the diagonal
gauge as follows. Assuming we put the reduced space to the form  $$
J= \pmatrix{0&k\cr -T&0\cr}.\eqn\tea$$  Then computing \coffee\
using this form of $J$,  we
obtain  (after solving for $v_2$ in terms of $v_1$) the
differential equation $$k\partial_z^2v_1+Tv_1=0.\eqn\vog$$
By identifying equations \vogue\ and \vog, we arrive at the form
of $T.$ The relation which connects $T$ to the free field
$\partial_z\phi$ is known as the Miura transformation [\Sok,\oog].

The analysis of [\birk] suggests that the phase space of
of N=1 Osp(1,2) current algebras can be reduced
to that of N=1 superconformal algebra. A similar observation has
been made in [\superkuramoto] in
which a quantum Hamiltonian reduction of super Osp(1,2) Kac-Moody
algebra has been performed.

This work is organized as follows. In the next section, the
results of [\birk] in which Polyakov's \lq\lq soldering
procedure" is used to show  how two-dimensional
N=1 superdiffeomorphisms can be obtained from N=1
Osp(1,2) gauge transformations is discussed using the
free-superfield realization of  N=1
Osp(1,2)  supercurrent algebra.
 In section 3,  we construct the free-
superfield realization of N=1 Osp(2,2) supercurrent algebra.
Using this representation, we impose a certain set of constraints
on the phase space of the supercurrent algebra and perform the
classical Hamiltonian reduction of the system.
The resulting phase space
is shown to have the N=2 superconformal symmetry. Section 4
contains a summary of our results and a  discussion of the quantum
theory.\chapter{ Classical N=1 superVirasoro
algebra} In this section the free-superfield representation of the
Osp(1,2) supercurrents [\superoog] is used to verify the results of
[\birk] in which the N=1 superconformal transformations were derived
as  restricted N=1 Osp(1,2) supergauge transformations. The  N=1
Osp(1,2) supercurrent algebra can be described in terms of one scalar
superfield $\Phi$, two pairs of superghosts $(B,C)$ and
$(\Psi^{\dagger},\Psi)$ \footnote *{The superghosts in components
are expressed as $B=b-\theta\beta$, $C=\gamma+\theta c$ where
$(b,c)$ are fermionic ghosts and $(\beta,\gamma)$ are bosonic.
Similarly $\Psi^{\dagger}=\alpha+\theta\psi^{\dagger}$ and
$\Psi=\psi+\theta\delta$ where the ghosts  $(\psi^{\dagger}, \psi)$
are fermionic and $(\alpha,\delta)$ are bosonic. We mention that it
is the ghosts $(\beta, \gamma)$ and $(\psi^{\dagger}, \psi)$ that
appear in the free-field realization of the bosonic Osp(1,2) current
algebra [\superoog].}of dimensions $(1/2, 0)$as   $$\eqalign{{\bf
J}^+=&kDC+2kCD\Phi+2 C\Psi\Psi^{\dagger}+C^2 B-kD\Psi\Psi,\cr {\bf
J}^-=& -B, \cr {\bf J}^0=&BC+kD\Phi+\Psi\Psi^{\dagger},\cr  {\bf
J}^{-{1\over2}}=&{1\over2}B\Psi-\Psi^{\dagger},\cr {\bf
J}^{+{1\over2}}=&kD\Psi-k\Psi D\Phi+C {\bf
J}^{-{1\over2}},\cr}\eqn\napalm$$ where $D=\partial_\theta
+\theta\partial_z$ and $\{0,+,-,-{1\over2},+{1\over2}\}$ are the
Osp(1,2) group indices. Imposing the partial gauge fixing of [\birk]
$${\bf J}^-={\bf J}^0={\bf J}^{+{1\over2}}=0, \qquad {\bf
J}^{-{1\over2}}=k, \quad {\bf J}^+=-{\bf {T}},\eqn\sowhat$$
gives the following relations among the free
superfields $$B=0, \qquad\Psi^{\dagger}=-k,
\qquad D\Phi=\Psi,\qquad C=-\partial_z\Phi.\eqn\saigon$$ These
relations when sustituted back in ${\bf T}$ give, $${\bf
{T}}=k(\partial_z\Phi D\Phi+\partial_z D\Phi).\eqn\horror$$
Obviously $\bf T$ is the Feign-Fuchs representation of a spin-3/2
superfield describing the phase space of N=1 superVirasoro algebra.
This can be easily demonstrated by computing the Poisson brackets of
the field components of ${\bf T}.$ Writing  $${\bf T}=G-\theta
T,\qquad \Phi=\phi+\theta\psi,\eqn\sophie$$ and using
$$\{D\Phi(z_1,\theta_1), \Phi(z_2,\theta_2) \}_{{\rm
P.B}}=-{1\over2k}\theta_{12}\delta(z_{12}),\eqn\dd$$ where
$$\theta_{12}=\theta_1-\theta_2, \qquad
z_{12}=z_1-z_2-\theta_1\theta_2,\eqn\lib$$the Poisson brackets of the
component fields
$$T=-k(\partial_z^2\phi+\partial_z\psi\psi+(\partial_z\phi)^2),
\qquad G=k(\partial_z\phi\psi+\partial_z\psi)\eqn\caren$$  can be
shown to satisfy the classical N=1 superVirasoro algebra,
$$\eqalign{\{T(z_1),T(z_2)\}_{{\rm P.B}}=&{k\over2}\partial_{z_2}^3
\delta(z_1-z_2)+ 2T(z_2)\partial_{z_2}\delta(z_1-z_2)+
\partial_{z_2} T(z_2)\delta(z_1-z_2),\cr \{T(z_1),G(z_2)\}_{{\rm
P.B}}=&\delta(z_1-z_2)\partial_{z_2}G(z_2)+
{3\over2}\partial_{z_2}\delta(z_1-z_2)G(z_2),\cr
\{G(z_1),G(z_2)\}_{{\rm P.B}}=&{k\over2}\partial_{z_2}^2
\delta(z_1-z_2)+{1\over2}\delta(z_1-z_2)T(z_2).\cr}\eqn\caroline$$
In the Hamiltonian reduction method, the partial gauge fixing
\sowhat\ is equivalent to the Drinfeld-Sokolov gauge, i.e, imposing
the constraints  $${\bf J}^-=0, \qquad {\bf
J}^{-{1\over2}}=k\eqn\okey$$ and then using the residual gauge
transformations (generated by the Borel super subgroup) to set ${\bf
J}^0$ and ${\bf J}^{+{1\over2}}$ to zero. However, in the diagonal
gauge one uses the residual gauge transformations to set ${\bf J}^+$
and ${\bf J}^{+{1\over2}}$ to zero. In this case the reduced space
is described by the free coordinate $D\Phi.$
The coordinates ${\bf T}$ and $D\Phi$ are connected by the super
Miura transformation.  This connection can be explained as follows.
Consider the system of linear differential equation
$$kDv-\pmatrix{{\bf J}^0&0&k\cr{\bf J}^+&-{\bf J}^0&{\bf
J}^{+{1\over2}}\cr {\bf J}^{+{1\over2}}&-k&0\cr}v=0,\qquad
v=\pmatrix{v_1\cr v_2\cr v_3}\eqn\cam$$ where $v$ is a vector. This
system of differential equations can be written into a higher order
gauge-invariant differential equation in terms of the component
$v_1$.  Computing \cam\ in the diagonal gauge
gives$$\eqalign{&kDv_1-kD\Phi v_1-kv_3=0,\cr &kDv_2+kD\Phi v_2=0,\cr
&kDv_3+kv_2=0.\cr }\eqn\colin$$ By solving these equation in terms
of $v_1$ we obtain  $$k\partial_z Dv_1-k(\partial_z
D\Phi+\partial_z\Phi D\Phi)v_1=0.\eqn\ram$$ In the Drinfeld-Sokolov
gauge one obtains from \cam\ the differential equations
$$\eqalign{&kDv_1-kv_3=0,\cr &kDv_2+{\bf T}v_1=0,\cr
&kDv_3+kv_2=0.\cr}\eqn\nes$$ Solving these equations in terms of
$v_1$ we get $$k\partial_z Dv_1-{\bf T}v_1=0.\eqn\ramziii$$
Comparing \ram\ and \ramziii\ we get the expression of ${\bf T}$.
For a discussion of the quantum hamiltonian reduction
of N=1 Osp(1,2) supercurrent algebra the reader is referred to
[\superkuramoto].  \chapter{Classical N=2 superVirasoro algebra} In
this section, motivated by the construction of [\bbirk] in which it
is shown that  N=2 superdiffeomorphisms are
restricted  N=1 supergauge transformations of a supergauge theory
with Osp(2,2) as a gauge group, we will consider in some details the
Hamiltonian reduction of the N=1 Osp(2,2) supercurrent algebra using
its free-superfield realization.
\section{Free- superfield
Realization of N=1 Osp(2,2) supercurrent algebra}   In this section
an explicit construction of the free superfields describing the phase
space of N=1 Osp(2,2) WZNW model is given. The orthosymplectic group
Osp(2,2) is generated by four bosonic generators $\{l_0, l_{-1},
l_{1}, l_u\} $ and four fermionic generators  $\{({l_{1\over2})}_1,
{(l_{1\over2})}_2, {(l_{-{1\over2}})}_1, {(l_{-{1\over2}})}_2\},$
which are represented as follows, $$l_0=\pmatrix{{1/2}&0&0&0\cr
0&-{1/2}&0&0\cr 0&0&0&0\cr  0&0&0&0\cr}\quad
l_{1}=\pmatrix{0&1&0&0\cr 0&0&0&0\cr 0&0&0&0\cr  0&0&0&0\cr}\quad
l_{-1}=\pmatrix{0&0&0&0\cr 1&0&0&0\cr 0&0&0&0\cr  0&0&0&0\cr}$$
$$l_u=\pmatrix{0&0&0&0\cr 0&0&0&0\cr 0&0&0&1\cr 0&0&- 1&0\cr}\quad
{(l_{{1\over2}})}_1=\pmatrix{0&0&1&0\cr 0&0&0&0\cr 0&-1&0&0\cr
0&0&0&0\cr}\quad {(l_{{1\over2}})}_2=\pmatrix{0&0&0&1\cr 0&0&0&0\cr
0&0&0&0\cr 0&-1&0&0\cr}$$
$${(l_{-{{1\over2}}})}_1=\pmatrix{0&0&0&0\cr 0&0&1&0\cr 1&0&0&0\cr
0&0&0&0\cr}\quad {(l_{-{{1\over2}}})}_2=\pmatrix{0&0&0&0\cr
0&0&0&1\cr  0&0&0&0\cr 1&0&0&0\cr}.\eqn\christ$$ A group element $g
\in \hbox{Osp(2,2)}$  can be represented by
$$g=\pmatrix{1&0&0&0\cr  C&1&\Psi_1&\Psi_2\cr \Psi_1&0&1&0\cr
\Psi_2&0&0&1\cr} \pmatrix{\lambda&0&0&0\cr 0&\lambda^{-1}&0&0\cr
0&0&G_1&G_2\cr  0&0&-G_2&G_1\cr} \pmatrix{1&F&\xi_1&\xi_2\cr
0&1&0&0\cr 0&-\xi_1&1&0\cr 0&-\xi_2&0&1\cr},\eqn\disaster$$ where
$\{C, \lambda, F, G_1, G_2\}$ are bosonic superfields
$(G_1^2+G_2^2=1)$ and $\{\Psi_1,  \Psi_2, \xi_1, \xi_2\}$ are
fermionic superfields.    The N=1 super Osp(2,2) WZNW action
[\Peterson,\FUCHS]  can then be expressed as
$$\eqalign{S\sim k\int &d^2zd^2\theta\Big(\bar D\Phi_0 D\Phi_0+
\bar D\Phi_1D\Phi_1 \cr &+\lambda^2(\bar DC-\Psi_1\bar D\Psi_1
-\Psi_2\bar D\Psi_2) (DF+D\xi_1\xi_1+D\xi_2\xi_2)\cr
&-2\lambda\bar D\Psi_1(G_1D\xi_1+G_2D\xi_2)
 -2\lambda\bar D\Psi_2(G_1D\xi_2-G_2D\xi_1),\cr}\eqn\nestor$$
where  the covariant derivatives $D$ and $\bar D$
are given in terms of the holomorphic and anti-holomorphic Grassmann
coordinates $\theta$ and $\bar \theta$ by
$$D\equiv{\partial\over\partial\theta}+
\theta{\partial\over\partial z}, \qquad
\bar D\equiv{\partial\over\partial\bar\theta}+\bar\theta{\partial
\over\partial\bar z}. \eqn\hi$$ The holomorphic supercurrents of
the super WZNW model are given by  $${\bf J}=kDgg^{-1}=
\pmatrix{{\bf J}^0&{\bf J}^{-}&{\bf J}_1^{-{1\over2}}&{\bf J}_2^{-{1\over2}}\cr
{\bf J}^+&-{\bf J}^0&{\bf J}_1^{+{1\over2}}&{\bf J}_2^{+{1\over2}}\cr
{\bf J}_1^{+{1\over2}}&-{\bf J}_1^{-{1\over2}}&0&-{\bf J}^u\cr
{\bf J}_2^{+{1\over2}}&-{\bf J}_2^{-{1\over2}}&{\bf J}^u&0\cr}.
\eqn\nice$$
If the following change of superfields is performed
$$\eqalign{B=-&k\lambda^2(DF+D\xi_1\xi_1+D\xi_2\xi_2),\cr
\Psi_1^{\dagger}=&{1\over2}B
\Psi_1-k\lambda(G_1D\xi_1+G_2D\xi_2),\cr
\Psi_2^{\dagger}=
&{1\over2}B\Psi_2-k\lambda(G_1D\xi_2-G_2D\xi_1),\cr}\eqn\trans$$
then the super WZNW action in \nestor\ becomes an
action of free superfields $$S\sim \int d^2zd^2\theta\Big(k\bar
D\Phi_0 D\Phi_0+ k\bar D\Phi_1 D\Phi_1+B\bar DC +
2\Psi_1^{\dagger}\bar D\Psi_1+
2\Psi_2^{\dagger}\bar D\Psi_1\Big).\eqn\hi$$

In terms of the free superfields the supercurrents can be
written as $$\eqalign{{\bf J}^+=&kDC+2kC
D\Phi_0+2C(\Psi_1\Psi_1^{\dagger}+\Psi_2\Psi_2^{\dagger}) +C^2B \cr
&-kD\Psi_1\Psi_1-kD\Psi_2\Psi_2-2k\Psi_1\Psi_2D\Phi_1,\cr {\bf
J}^-=&-B,\cr {\bf J}^0=&BC+kD\Phi_0+\Psi_1\Psi_1^{\dagger}+
\Psi_2\Psi_2^{\dagger},\cr {\bf
J}^u=&-\Psi_1\Psi_2^{\dagger}+\Psi_2\Psi_1^{\dagger}-kD\Phi_1,\cr
{\bf J}_1^{-{1\over2}}=&{1\over2}B\Psi_1-\Psi_1^{\dagger},\cr {\bf
J}_2^{-{1\over2}}=&{1\over2}B\Psi_2-\Psi_2^{\dagger},\cr {\bf
J}_1^{+{1\over2}}=&kD\Psi_1-k\Psi_1D\Phi_0 +C{\bf J}^{-{1\over2}}_1
-\Psi_1\Psi_2\Psi_2^{\dagger} +k\Psi_2D\Phi_1,\cr {\bf
J}_2^{+{1\over2}}=&kD\Psi_2-k\Psi_2D\Phi_0 +C {\bf J}^{-{1\over2}}_2
+\Psi_1\Psi_2\Psi_1^{\dagger} -k\Psi_1D\Phi_1.\cr}\eqn\peace$$ where
$$G_1={\hbox {sin}}\ \Phi_1,\qquad \lambda=\exp {(\Phi_0)}.$$

\section{Hamiltonian Reduction}
Having constructed the phase space of the N=1 super Osp(2,2) in
terms of free superfields, we now proceed to show that this phase
space can be reduced to that of the
N=2 superVirasoro algebra. The constraints we impose are the
following,  $${\bf J}^-=0, \quad {\bf
J}_1^{-{1\over2}}=0,\quad {\bf
J}_2^{-{1\over2}}=k.\eqn\constraint$$  An element of the space of
the residual symmetry that preserves these constraints takes values
in the Borel super subalgebra generated by $\{l^{-1},
l_1^{-{1\over2}}, l_2^{-{1\over2}}\}$ and is given by
$$U=\pmatrix{1&0&0&0\cr
A&1&\epsilon_1&\epsilon_2\cr\epsilon_1&0&1&0\cr
\epsilon_2&0&0&1\cr}.\eqn\borel$$ Under the finite residual gauge
transformation $$g\rightarrow Ug,\eqn\transformation$$ the
supercurrents transform as $$\eqalign{{\bf J}\rightarrow
&\pmatrix{1&0&0&0\cr
A&1&-\epsilon_1&-\epsilon_2\cr-\epsilon_1&0&1&0\cr
-\epsilon_2&0&0&1\cr}{\bf J}\pmatrix{1&0&0&0\cr
-A&1&-\epsilon_1&-\epsilon_2\cr-\epsilon_1&0&1&0\cr
-\epsilon_2&0&0&1\cr}\cr &+k\pmatrix{0&0&0&0\cr
DA-D\epsilon_2\epsilon_2-D\epsilon_1\epsilon_1&0&D\epsilon_1&D
\epsilon_2\cr D\epsilon_1&0&0&0\cr
D\epsilon_2&0&0&0\cr}.\cr}\eqn\lax$$ If we select the gauge
parameters as (Drinfeld-Sokolov gauge), $$\epsilon_2={1\over k}{\bf
J}^0,\qquad \epsilon_1=-{1\over k}{\bf J}^u,\qquad A=-{1\over k}{\bf
J}_2^{+{1\over2}}-{1\over k}D{\bf J}^0,\eqn\residual$$ then \lax\  can be used
to bring ${\bf  J}$ into the following form $${\bf
J}_{DS}=\pmatrix{0&0&0&k\cr X&0&Y&0\cr Y&0&0&0\cr
0&-k&0&0\cr},\eqn\safia$$ where $X$ and $Y$,  the coordinates of
reduced phase space, are given by $$\eqalign{X=&{\bf
J}^++{2\over k}{\bf J}_1^{+{1\over2}}{\bf J}^u-{2\over k}{\bf
J}_2^{+{1\over2}}{\bf J}^0-D{\bf J}_2^{+{1\over2}}-\partial_z{\bf
J}^0-{1\over k}D{\bf J}^0{\bf J}^0-{1\over k}D{\bf J}^u{\bf J}^u,
\cr Y=&{\bf
J}_1^{+{1\over2}}-{1\over k}{\bf J}^0{\bf J}^u-D{\bf
J}^u.\cr}\eqn\rafic$$ Substituting the expressions of the
supercurrents in \rafic\ and using the constraints \constraint, we
get  $$\eqalign{X=&-k(\partial_z\Phi_1D\Phi_1+
\partial_z\Phi_0D\Phi_0+ +\partial_zD\Phi_0), \cr
Y=&k(\partial_z\Phi_1+D\Phi_0 D\Phi_1).\cr}\eqn\hiam$$  In terms of
the components, $$\eqalign{&\Phi_0=\phi_0+\theta\psi_0,\qquad
\Phi_1=\phi_1+\theta\psi_1,\cr &X=-(G_1-\theta T), \qquad Y=U+\theta
G_2,\cr}\eqn\components$$ equation \hiam\ gives,
$$\eqalign{T=&-k(\partial_z\psi_1\psi_1+\partial_z\psi_0\psi_0
+\partial_z\phi_1\partial_z\phi_1+\partial_z\phi_0
\partial_z\phi_0+\partial_z^2\phi_0),\cr
G_1=&k(\partial_z\psi_0+\psi_1\partial_z\phi_1+
\psi_0\partial_z\phi_0),\cr
G_2=&k(\partial_z\psi_1-\psi_0\partial_z\phi_1+
\psi_1\partial_z\phi_0),\cr
U=&k(\partial\phi_1+\psi_0\psi_1).\cr}\eqn\lebanon$$ Using
$$\eqalign{\{D\Phi_0(z_1,\theta_1), \Phi_0(z_2,\theta_2)
\}_{{\rm P.B}}=&-{1\over2k}\theta_{12}\delta(z_{12}),\cr
\{D\Phi_1(z_1,\theta_1), \Phi_1(z_2,\theta_2)
\}_{{\rm
P.B}}=&-{1\over2k}\theta_{12}\delta(z_{12}),\cr}\eqn\poisson$$
the Poisson brackets of the coordinates $\{T,G_1,G_2,U\}$ can
be easily shown to satisfy the classical N=2 superconformal
algebra,$$\eqalign{\{T(z_1),T(z_2)\}_{{\rm
P.B}}=&{k\over2}\partial_{z_2}^3 \delta(z_1-z_2)+
2T(z_2)\partial_{z_2}\delta(z_1-z_2)\cr &
+\partial_{z_2}T(z_2)\delta(z_1-z_2),\cr \{T(z_1),G_1(z_2)\}_{{\rm
P.B}}=&\delta(z_1-z_2)\partial_{z_2}G_1(z_2)+
{3\over2}\partial_{z_2}\delta(z_1-z_2)G_1(z_2),\cr
\{T(z_1),G_2(z_2)\}_{{\rm
P.B}}=&\delta(z_1-z_2)\partial_{z_2}G_2(z_2)+
{3\over2}\partial_{z_2}\delta(z_1-z_2)G_2(z_2),\cr
\{T(z_1),U(z_2)\}_{{\rm P.B}}=&\delta(z_1-z_2)U(z_2)+
\partial_{z_2}\delta(z_1-z_2)U(z_2),\cr \{G_1(z_1),G_1(z_2)\}_{{\rm
P.B}}=&{k\over2}\partial_{z_2}^2
\delta(z_1-z_2)+{1\over2}\delta(z_1-z_2)T(z_2),\cr
\{G_2(z_1),G_2(z_2)\}_{{\rm P.B}}=&{k\over2}\partial_{z_2}^2
\delta(z_1-z_2)+{1\over2}\delta(z_1-z_2)T(z_2),\cr
\{U(z_1),G_1(z_2)\}_{{\rm
P.B}}=&{1\over2}\delta(z_1-z_2)G_2(z_2),\cr
\{U(z_1),G_2(z_2)\}_{{\rm
P.B}}=&-{1\over2}\delta(z_1-z_2)G_1(z_2),\cr
\{G_1(z_1),G_2(z_2)\}_{{\rm P.B}}=&
-{1\over2}\delta(z_1-z_2)\partial_{z_2}U(z_2)
-\partial_{z_2}\delta(z_1-z_2)U(z_2),\cr \{U(z_1),U(z_2)\}_{{\rm
P.B}}=&-{k\over2}\partial_{z_2}\delta(z_1-z_2).\cr}\eqn\carolyn$$
The diagonal gauge is given by
$$\epsilon_1=-\Psi_1,\qquad\epsilon_2=-\Psi_2,\qquad
A=-C.\eqn\diagonal$$ In this gauge $\bf J$ takes  the following form
$${\bf J}_{diagonal}=\pmatrix{kD\Phi_0&0&0&k\cr 0&-kD\Phi_0&0&0\cr
0&0&0&kD\Phi_1\cr 0&-k&-kD\Phi_1&0\cr}.\eqn\croatia$$ The
coordinates $\{X, Y\}$ are connected to the coordinates $\{D\Phi_0,
D\Phi_1\}$ by the super Miura transformation.  This connection can
be explained as in the previous cases as follows.  Consider the
system of linear differential equation $$(kD-{\bf J})v=0,\qquad
v=\pmatrix{v_1\cr v_2\cr v_3\cr v_4}.\eqn\camomile$$  Computing
\camomile\ in the diagonal gauge gives $$\eqalign{&kDv_1-kD\Phi_0
v_1-kv_4=0,\cr &kDv_2+kD\Phi_0 v_2=0,\cr &kDv_3-kD\Phi_1 v_4=0,\cr
&kDv_4+kv_2+kD\Phi_1v_3=0.\cr}\eqn\nest$$ By solving these equation
in terms of $v_1$, the following gauge-invariant differential
equation is obtained,
$$k\Big(\partial_z
D-(\partial_z\Phi_0v_1+D\Phi_0\partial_z\Phi_0)+
k(\partial_z\Phi_1+D\Phi_0D\Phi_1){1\over D}
(\partial_z\Phi_1+D\Phi_0D\Phi_1)\Big)v_1=0.\eqn\ramzi$$ In the
Drinfeld-Sokolov gauge, \camomile\ becomes
$$\eqalign{&kDv_1-kv_4=0,\cr &kDv_2-X v_1-Yv_3=0,\cr &kDv_3-Y
v_1=0,\cr &kDv_4+kv_2=0.\cr}\eqn\neste$$ Similarly, by solving these
equations in terms of  $v_1$ we get
$$k\partial_z
Dv_1+Xv_1+Y{1\over D}(Yv_1)=0.\eqn\ramzii$$
Comparing \ramzi\ and \ramzii\ we get the expressions of $X$ and $Y$
in \hiam.

\chapter{Discussions}
In this letter, the Hamiltonian reduction of the SL(2,R) current
algebra is reviewed using its free-field realization. We also have
demonstrated the equivalence of this formalism to Polyakov \lq\lq
Soldering procedure". The results of [\birk] were also verified
using the free-superfield realization of of the N=1 Osp(1,2)
supercurrent algebra. Motivated by the results of [\bbirk] in which
the Polyakov method has been employed to extract the N=2
superconformal transformations from the N=1 Osp(2,2) supergauge
transformation, we have considered the classical Hamiltonian
reduction of the N=1 Osp(2,2) supercurrent algebra using its
free-superfield realization. The reduced phase space was shown to
exhibit the classical N=2 superconformal algebra.
Our construction can be extended to the cases of
N=1 Osp$(n,2)$ $(n>2)$ supercurrent algebras. We expect that these
algebras will reduce to the $O(n)$-extended superconformal
algebras found by Bershadsky and Knizhnik [\B].

Finally we give some comments on the quantum theory.
The quantum WZNW model with a group $G$ is a
superconformal and super $G$ Kac-Moody invariant two dimensional
field theory. The super stress-energy tensor ${\cal T}$
constructed from the super Kac-Moody currents via the super Sugawara
construction [\Peterson,\FUCHS] together with the supercurrents
satisfy the following operator product expansion $$\eqalign{&{\cal
T}(z_1,\theta_1){\cal T}(z_2,\theta_2) \sim
{c\over6}{1\over z^3_{12}}+{3\over2}{\theta_{12}\over
z_{12}^2}{\cal T}(z_2,\theta_2)+{1\over 2 z_{12}}D{\cal
T}(z_2,\theta_2)+{\theta_{12}\over z_{12}}\partial{\cal
T}(z_2,\theta_2),\cr &{\cal T}(z_1,\theta_1)
{\bf J}^{(a)}(z_2,\theta_2)\sim {1\over2}{\theta_{12}\over
z^2_{12}}{\bf J}^{a}(z_2,\theta_2)+
{1\over2}{D_2{\bf J}^{a}(z_2,\theta_2)\over z_{12}}
+{\theta_{12}\partial_{z_2}{\bf J}^{a}(z_2,\theta_2)\over
z_{12}},\cr &{\bf J}^{a}(z_1,\theta_1){\bf J}^{b}(z_2,\theta_2)\sim
{{k/2}\eta^{ab}\over z_{12}}
+{\theta_{12}f^{ab}{}_{c}{\bf J}^{c}(z_2,\theta_2)\over
z_{12}},\cr}\eqn\liban$$ where $\eta^{ab}$ and $f^{ab}{}_{c}$ are
the metric and structure constant of the Lie algebra $\cal G$ of
$G$. The central charge of the superVirasor algebra is given by
$$c={3\over2}\Big(1-{2C_{\hbox{adj}}\over k}\Big)dim{\cal
G},\eqn\susy$$ where $C_{\hbox{adj}}$ is the dual Coexeter number of
the group $G$ and $dim$ is the dimension of the group $G$ (in the
case when $G$ is a supergroup the $dim$ is replaced by the $sdim$,
where $sdim$ is the $dim$ of the bosonic generators minus the $dim$
of the fermionic generators of the group $G$). The second equation
in \liban\ means that all the supercurrents have the dimension $1/2$
with respect to the super stress-energy tensor ${\cal T}.$

In the analysis of [\oog] where the quantum Hamitonian reduction of
the SL(2,R) bosonic current algebra is considered, the stress
tensor has to be modified in order that the constraints
are consistent with conformal invariance.
However, in the case of the
Hamiltonian reduction of the bosonic Osp(1,2)  current algebra
[\superoog], after imposing the constraint $J^-=1$ and modifiying
the stress tensor as in the case of SL(2,R), additional
Majorana fermion has to be introduced for the closure of the algebra
of constraints. Similarly, in the Hamiltonian reduction of Osp(2,2)
current algebra two Majorana fermions have to be introduced in order
that the constraints form a closed algebra.

In [\superkuramoto] where the quantum Hamiltonian reduction of the
Osp(1,2) supercurrent algebra is considered, an appropriate super
stress-energy tensor is constructed by modifying the Sugawara form of
the super stress-energy tensor as
$${\cal T}'={\cal T}-\partial_z{\bf J}^{0}.\eqn\tania$$
In this construction no additional structure has to be introduced as
the superspace
constraints are consistent and form a closed  algebra. This is
because the superspace current ${\bf J}^{(-{1\over2})}$ is of
dimension $0$ with respect to ${\cal T}'.$

The same situation obviously arises in the quantum Hamiltonian
reduction of Osp(2,2) supercurrent algebra. After modifying the
super stress-energy tensor as in \tania, the supercurrents
${\bf J}_1^{(-{1\over2})}$ and ${\bf J}_2^{(-{1\over2})}$
are both of dimension $0$ and the algebra of constraints closes
without introducing extra fermions. Then following the standard
procedure of quantizing a system with constraints, one introduces
three superspace ghost systems (corresponding to the three
constraints), $(b(z, \theta), c(z, \theta)),$ $(\eta (z,\theta),
\zeta (z,\theta))$  and $(\eta' (z,\theta), \zeta'(z,\theta))$
 with dimensions $(-{1\over2}, 1),$ $(0,{1\over2}),$ and
$(0,{1\over2})$ respectively.
In components, these superghosts are
given as $$\eqalign{b(z, \theta)=b_1+\theta b_2,& \qquad c(z,
\theta)=c_1+\theta c_2,\cr \eta (z, \theta)=\eta_1
+\theta\eta_2, &\qquad \zeta(z, \theta)=\zeta_1+\theta\zeta_2, \cr
\eta' (z, \theta)=\eta'_1
+\theta\eta'_2, &\qquad \zeta'(z,
\theta)=\zeta'_1+\theta\zeta'_2.\cr}\eqn\friedan$$ The
ghosts $(b_2, c_1),$ $(\eta_1, \zeta_2)$ and $(\eta'_1,
\zeta'_2)$ are all anticommuting and have the  dimension $(0,1)$
while  the pairs $(b_1,c_2),$ $(\eta_2,\zeta_1)$ and
$(\eta'_2,\zeta'_1)$ are all commuting and have the dimensions of
$(-{1\over2}, {3\over2}),$ $({1\over2},{1\over2})$ and
$({1\over2},{1\over2})$ respectively. The total super strees-energy
tensor ${\bf T}^{\hbox {total}}$ of the  system is the sum of the
modified super stress energy tensor and that of the  superghosts. The
superghosts contribute $+3$ to the superconformal anomaly.
Thus the total central charge of the reduced theory is given by
$$c^{\hbox{tot}}=-6k+3.\eqn\re$$ In order to complete the quantum
Hamiltonian reduction, the reduced Hilbert space must be obtained.
One has to construct  the  BRST operator and study its cohomology
following [\oog,\superoog,\superkuramoto].  Then this cohomology
must be shown to be isomorphic to an irreducible representation
space of the N=2 superVirasoro algebra with central charge
$c=3p/(p+2)$ (where $k=1/(p+2)$). We hope to report on this in a
future publication.

\centerline{ACKNOWLEDGEMENT}
I would like to thank T. Kuramoto for useful
discussions.
\refout
\end